\documentclass[letter]{jpsj2}
\usepackage{graphicx}
\usepackage{dcolumn}
\usepackage{bm}
\usepackage{amsmath}
\usepackage{amssymb}

\begin{document}
\title{Scale-free networks with self-growing weight}
\author{Takuma Tanaka$^1$\thanks{E-mail address:
ttakuma@mbs.med.kyoto-u.ac.jp}, Toshio Aoyagi$^{2,3}$
}
\inst{%
$^1$Department of Morphological Brain Science, Graduate School
of Medicine, Kyoto University, Japan, 606-8501\\
$^2$Department of Applied Analysis and Complex Dynamical Systems,
Graduate School of Informatics, Kyoto University, Japan\\
$^3$CREST, JST
}

\date{\today}
\abst{%
We present a novel type of weighted scale-free network model, in which
the weight grows independently of the attachment of new nodes.
The evolution of this network is thus determined not only by the
preferential attachment of new nodes to existing nodes but also by
self-growing weight of existing links based on a simple weight-driven
rule. 
This model is analytically tractable, so that the various statistical properties, such as the
 distribution of weight, can be derived. 
Finally, we found that some type of social networks is well 
described by this model.
}

\kword{%
Weighted scale-free network, power-law distribution, complex network,
social network, natural language
}

\newcommand{\changedtext}{}


\maketitle

\newpage

The past decade has witnessed an explosive advance in the understanding of  the
network structures emerging in many fields, such as networks of
protein-protein interaction \cite{Jeong:2001}, the WWW \cite{Albert:1999},
and the Internet \cite{Willinger:2002}.
The most remarkable salient topological feature of these
networks is scale-freeness, that is, the power-law degree distribution. 
Theoretical studies have revealed two essential mechanisms
to generate such scale-free networks \cite{Albert:2002}: growth through the
continuous
addition of new nodes and preferential attachment of new nodes to the
existing nodes with higher connectivity.
Thus, we can say that the
scale-free properties of the network can be
successfully explained by these two mechanisms,
though some alternative models with the use of quenched disorder fitness distributions have been proposed
for generating static scale-free networks \cite{Caldarelli:2002}.

In the above cases, we take into account only the topological
network structure, in which the links between nodes are either
present or not.
However, beyond such purely topological structures, the
interaction strength through the link, i.e., the weight of the link plays a
crucial role in real-world networks, particularly when we need to
consider some dynamical systems on the network.
In the network of airports, for example, the number of
passengers traveling between two airports can be regarded as the weight
of the link connecting these airports \cite{Barrat:2004b}.
Similarly, coauthorship \cite{Newman:2001a,Barrat:2004b} and natural language \cite{FerreriCancho:2001} are known to be weighted scale-free networks.
In these weighted scale-free networks, it
is reported that not only the distribution of the degree of the nodes
but also that of the weights of the links obeys a power law.
Hence,
we will attempt to understand how such weighted scale-free networks
appear, in other words, whether there exist some specific mechanisms
underlying the power law of distribution of link weights.

To account for these power laws, several models have already been
proposed recently.
Most of the previous studies \cite{Barrat:2004a,Yook:2001,Zheng:2003} modeled weighted
networks assuming that the weight once assigned either remains unaltered
or is readjusted only when new nodes are added.
Wang and Zhang \cite{Wang:2004}
reported a model network which grows through preferential
attachment.
The growth is then determined by the fitness and the degree of
nodes independently of the weights of links.
In any case, the weight of
a link does not grow by itself independently of the attachment of new
nodes and degree of the node.
It seems that some real-world networks can be explained by these
models.
However, in many other real-world networks, the weight of a link
can grow spontaneously through a certain weight-driven mechanism.
\changedtext{
For example, in the coauthorship network of the researchers, a node
corresponds to a researcher and two nodes are connected by a link
with a weight.
The value of this weight is defined by the number of
papers on which the two corresponding persons collaborated.
In this case, when two persons collaborate again in
another paper, the weight of the link increases without making new edges.
An excellent researcher has collaborated with many other researchers on many
papers, which implies the sum of the weights of the links connecting to
the corresponding node (which we call the strength of the node) is very large.
In addition, such an important researcher tends to write many papers.
}
This means that links connecting to stronger nodes tend to increase their weight more rapidly, which is a characteristic
of weight-driven preferential attachment.
In this paper, we propose
an analytically tractable model of weighted complex networks which grow
through the preferential attachment driven by the strengths of the
nodes.
We show that power-law distributions of the degree, weight, and strength can be
derived theoretically and that these results are confirmed numerically.
\changedtext{
Moreover, we demonstrate that the networks of coauthorship
and e-mail can be well explained by this model.
}

Before introducing our model, we define some measures to
characterize weighted networks.
First, the connectivity of a network can be
expressed by an adjacency matrix $a_{ij}$, whose elements take the
value 1 if the node $i$ is connected to the node $j$ and 0 otherwise.
The degree of node $i$ is then defined by $k_i =\sum_{j=1} ^N a_{ij},$
where $N$ is the total number of nodes.
In addition, the weight of the link between nodes $i$ and $j$ is
denoted by $w_{ij}$.
Let us define the strength of node $i$, $s_i$ as
$s_i =\sum_{j=1} ^{N} a_{ij} w_{ij},$
which is the sum of the weight of all the links connecting to 
node $i$.
In this model, we assume that the links are undirected,
so that the adjacency matrix $a_{ij}$ and the weight matrix $w_{ij}$ are
symmetric.

\begin{figure}
\resizebox{80mm}{!}{\includegraphics{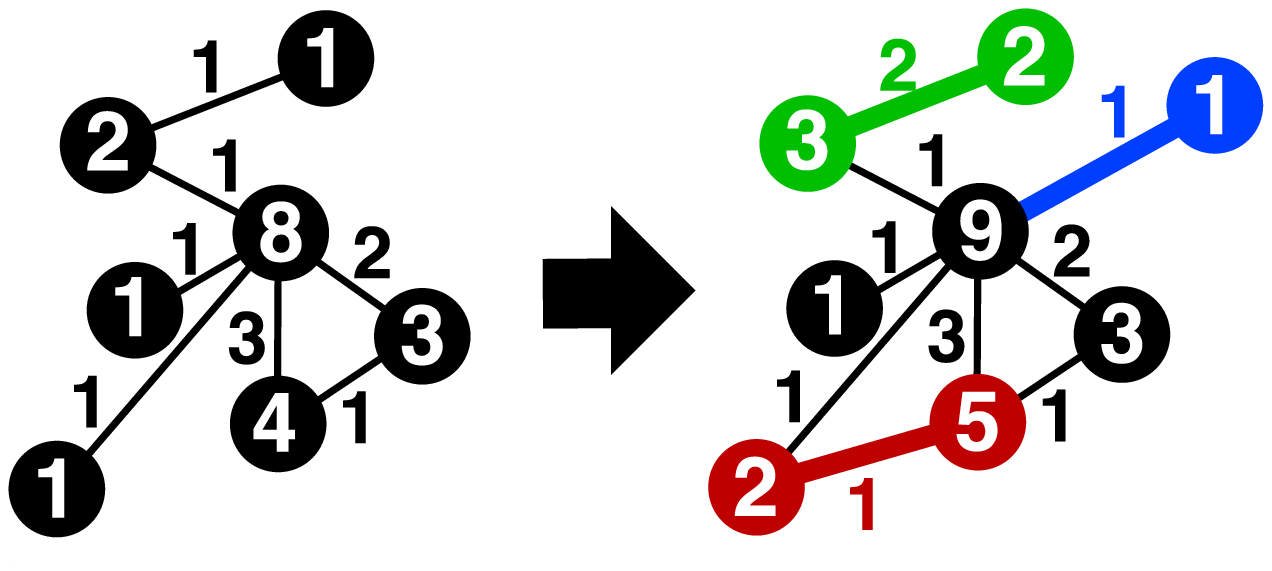}}
\caption{\label{schema}}
\end{figure}

We present a set of rules for generating the network as follows (Fig.~\ref{schema}).
The network initially starts with a single node.
Rule 1: at each time step, a new node is added to the
network and a connection is made to one existing node $i$, where the
probability that the node $i$ is chosen is proportional to the strength,
i.e. $s_i/\sum_{j=1}^N s_j$ (strength-driven preferential attachment).
The weight of this new link is then set to unity. Rule 2: at each time step, $ct$ pairs of the existing nodes are selected with
the probability proportional to their strengths, i.e. $s_i s_j
/\left(\sum_{k=1}^N s_k\right)^2$.
If these two nodes are not connected, they are connected by
a link with the weight equal to unity. If they are already connected,
the weight of the corresponding link between them is incremented by one.
This rule can be regarded as a generalization of the rule in the word
web growth \cite{Dorogovtsev:2001}.
Note that the total number of nodes is equal to the
time $t$ and each node can be labeled by the time $u$ when the node is
added.
Both the creation of new links and the changes in the weight of
existing links increase the strength of nodes, and the strength of a node
increases on average by approximately $2c$ at each time step.
In the case of the movie, for example, this implies 
that an actor/actress plays
together with, on average, $2c$ actors/actresses at each time step.

\begin{figure}
\resizebox{80mm}{!}{\includegraphics{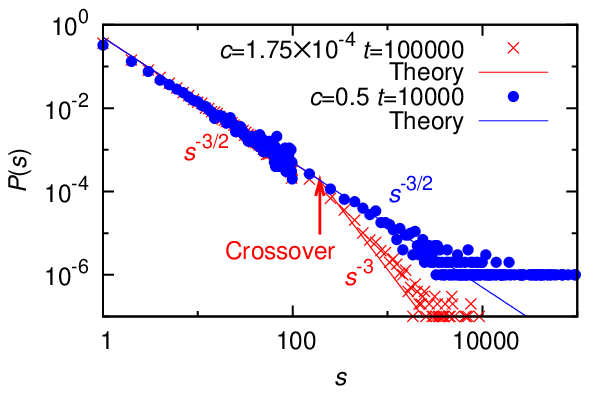}}
\caption{\label{fig1}
}
\end{figure}

To analytically obtain the the statistical properties of the network
generated by the above algorithm, we use a continuous approximation.
Now, let us denote the averaged strength of the node at time $t$ by $s(u, t)$,
where $u$ is the time at which this node was added to the network.
In the same way as \cite{Dorogovtsev:2001}, the time evolution of $s(u, t)$
is described by the equation
\begin{equation}
\tfrac{\partial s(u,t)}{\partial t}=(1+2ct)\tfrac{s(u,t)}{\int_0 ^t dv\,
s(v,t)} \label{strengthdef}
\end{equation}
with boundary condition $s(t,t)=1$.
Substituting the normalized condition $\int_0 ^t dv\,
s(v,t)=2t+ct^2$ into eq.~\ref{strengthdef}, we obtain the solution
\begin{equation}
s(u,t)=\sqrt{t (2+ct)^3/\left[u (2+cu)^3\right]}. \label{strength}
\end{equation}
For $cu\ll 1$, the distribution of the strength takes the form
$P(s)\approx\tfrac{(2+ct)^3}{4} s^{-3}$ because $s(u,t)$ is approximated by
 $\sqrt{t (2+ct)^3 u^{-1}2^{-3}}$.
For $cu\gg 1$, the approximation $s(u,t) \approx \sqrt{t (2+ct)^3 u^{-4} c^{-3}} $ gives $P(s) \approx \tfrac{1}{2}\left(\tfrac{t(2+ct)^3}{c^3 t^4}\right)^{1/4}
s^{-3/2} \approx 1/(2s^{3/2})$. 
Fig.~\ref{fig1} shows the comparison of distribution of strength between
theoretical and numerical results, in which the exponents obtained in the
simulations agree well with theoretical ones.

Similarly,
as a continuous version of the adjacency matrix $a_{ij}$, let us consider the
averaged connectivity of the nodes at time $t$, $a(u_1,u_2,t)$, where two
nodes at each end of the link are added at time $u_1$ and $u_2$.
The connectivity $a(u_1,u_2,t)$ satisfies the differential equation
\[\tfrac{\partial a(u_1 ,u_2 ,t)}{\partial t} = 2ct\tfrac{s(u_1 ,t)s(u_2 ,t)}{\left(\int_0 ^t dv\, s(v,t)\right)^2 } [1-a(u_1 ,u_2 ,t)],\]
which has a general solution
\begin{eqnarray}
&&a(u_1 ,u_2 ,t) \nonumber \\
&=&1-\exp \left(-\tfrac{(2+ct)^2 }{\sqrt{u_1 u_2 (2+cu_1 )^3 (2+cu_2 )^3} }\right) F(u_1 ,u_2 ),
\label{connectivity}
\end{eqnarray}
where $F(u_1 ,u_2 )$ is an arbitrary function.
Although the boundary condition
\begin{eqnarray*}
a(t ,u ,t)=a(u, t, t) &=& \tfrac{s(u,t)}{\int_0 ^t dv\, s(v,t)} 
= \sqrt{\tfrac{2+ct }{t u  (2+cu )^3 }} 
\end{eqnarray*}
cannot be satisfied, we set $F(u_1 ,u_2 )=1$ and Taylor expand to obtain
\begin{eqnarray*}
a(t ,u ,t) = a(u ,t ,t) &=& \sqrt{\tfrac{2+ct }{t u (2+cu )^3 }}+\cdots
\approx \tfrac{s(u,t)}{\int_0 ^t dv\, s(v,t)}.
\end{eqnarray*}
Hence, eq.~\ref{connectivity} is an approximate solution of
connectivity if $F(u_1 ,u_2)=1$.

\begin{figure}
\resizebox{80mm}{!}{\includegraphics{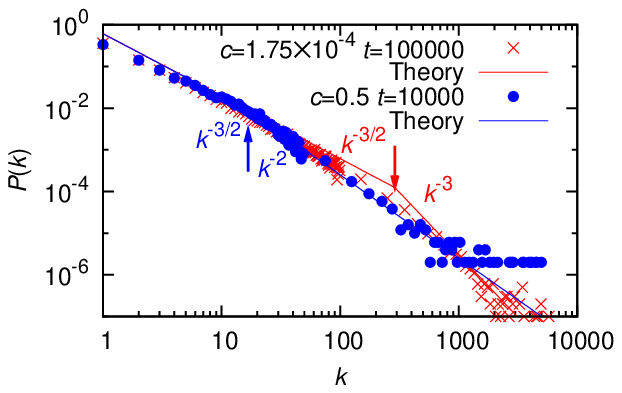}}
\caption{\label{fig3}
}
\end{figure}

The average degree of the node born at time $u$ is given by
\begin{eqnarray}
k(u,t)&=&\int_{0} ^t dv\, a(u ,v, t) \nonumber \\
&\approx& \displaystyle \int_{0} ^{2/c} dv \left[1-\exp \left( -\tfrac{(2+ct)^2 }{\sqrt{uv (2+cu)^3 2^3 }}\right) \right] \nonumber \\
&&+ \int_{2/c} ^t dv \left[ 1-\exp \left( -\tfrac{(2+ct)^2 }{\sqrt{uv (2+cu)^3 (cv)^3 }} \right) \right] \nonumber \\
&=& t - \int_0 ^{2/c} dv\, \exp \left(-A/\sqrt{8v}\right) \nonumber \\
&& \phantom{t}- \int_{2/c} ^t dv \exp \left[ -A /\left(\sqrt{c^3 } v^2 \right) \right] \nonumber\\
&=& t-\Bigl[\left(\tfrac{2}{c}-\tfrac{A}{2\sqrt{c}}\right)
\exp \left(-\tfrac{\sqrt{c}A}{4}\right)\nonumber\\
&&\phantom{t-\Bigl[}+\tfrac{A^2}{8}\Gamma \left(0,\tfrac{\sqrt{c}A}{4}\right)\Bigr] \nonumber\\
&&\phantom{t}-\left[\exp\left(-\tfrac{A}{\sqrt{c^3}v^2}\right)+\sqrt{\tfrac{\pi A}{\sqrt{c^3}}}\mathrm{erf}\left(\sqrt{\tfrac{A}{\sqrt{c^3}}}\tfrac{1}{v}\right)\right]^{t}_{2/c}\nonumber\\
&\approx& \begin{cases}
\sqrt{\pi A/\sqrt{c^3}} & (\sqrt{c}A\gg 1)\\
3A/(2\sqrt{c}) & (\sqrt{c}A\ll 1),
\end{cases} \label{degree2}
\end{eqnarray}
where $A =(2+ct)^2 /\sqrt{u(2+cu)^3 } $ and $\Gamma (a,b)$ is the incomplete gamma function.
If $c$ is small and $\sqrt{c}A\ll 1$ holds for all $u$, the degree distribution takes the form
$P(k)\approx \sqrt{3/8}k^{-3/2}$ for $cu\gg 1$
and
$P(k)\approx 9(ct)^3/(16k^3)$ for $cu\ll 1$.
If $c$ is large, assuming $cu\gg 1$ we obtain two different degree
distributions: 
$P(k)\approx \sqrt{\pi/c}/k^{2}$ for $\sqrt{c}A\gg 1$
and
$P(k)\approx \sqrt{3/8}k^{-3/2}$ for $\sqrt{c}A\ll 1$ (Fig.~\ref{fig3}).

From Eqs.~\ref{strength} and \ref{degree2}, we find the relationship
between the degree and the strength. 
The degree $k$ is proportional to strength $s$ for $\sqrt{c}A\ll 1$,
whereas $k \approx \sqrt{\pi /c} s^{1/2}$ holds for $\sqrt{c}A\gg 1$
(data not shown).
The linear relationship for $\sqrt{c}A\ll 1$ comes from the fact that 
the weights of almost all links between `young' nodes equal unity.

\begin{figure}
\resizebox{80mm}{!}{\includegraphics{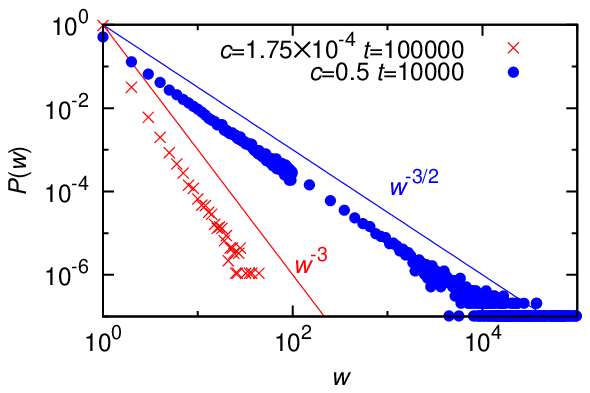}}
\caption{\label{fig5} 
}
\end{figure}

As is the case of the adjacency matrix,
we can define the continuous version of the weight matrix $w_{ij}$, $w(u_1 ,u_2 ,t)$,
whose dynamics are governed by the differential equation
\[\tfrac{\partial w(u_1 ,u_2 ,t)}{\partial t} = 2ct\tfrac{s(u_1 ,t)s(u_2 ,t)}{\left(\int_0 ^t dv\, s(v,t)\right)^2 }.\]
The solution is given by
\[w(u_1 ,u_2 ,t) = (2+ct)^2 /\sqrt{u_1 u_2 (2+cu_1 )^3 (2+cu_2 )^3} .\]
Note that the relationship $\int_0 ^{t} dv\, w(u ,v ,t) = s(u,t)$ is satisfied.
The distribution of weights of all links in the network is given by
\begin{eqnarray*}
P(w) &=& C \int_0 ^t du_1 \left| \frac{\partial (u_1 ,u_2)}
{\partial (u_1 ,w)}\right| a(u_1 ,u_2 ,t)  \\
 &\approx& C \int_{2/c} ^t du_1 \frac{du_2}{dw}[1-\exp (-w)],
\end{eqnarray*}
where $C$ is the normalization constant.
Using $P(w)\sim \tfrac{du_2}{dw}$ for large $w$, we obtain
$P(w)\sim w^{-3/2}$ if $cu_2\gg 1$ and $P(w)\sim w^{-3}$ if
$cu_2\ll 1$ (Fig.~\ref{fig5}).
In addition, it is often observed that the average weight scales with the
degrees of the nodes as $\langle w_{ij} \rangle =(k_i k_j)^\theta$
\cite{Barrat:2004b}.
We obtain $\theta = 2$ for the first regime of eq.~\ref{degree2}
if $cu_i\gg 1$ and $cu_j\gg 1$, and $\theta = 1$ if $\sqrt{c}A\ll
1$ for all $u$ (data not shown).

\begin{figure}
\resizebox{80mm}{!}{\includegraphics{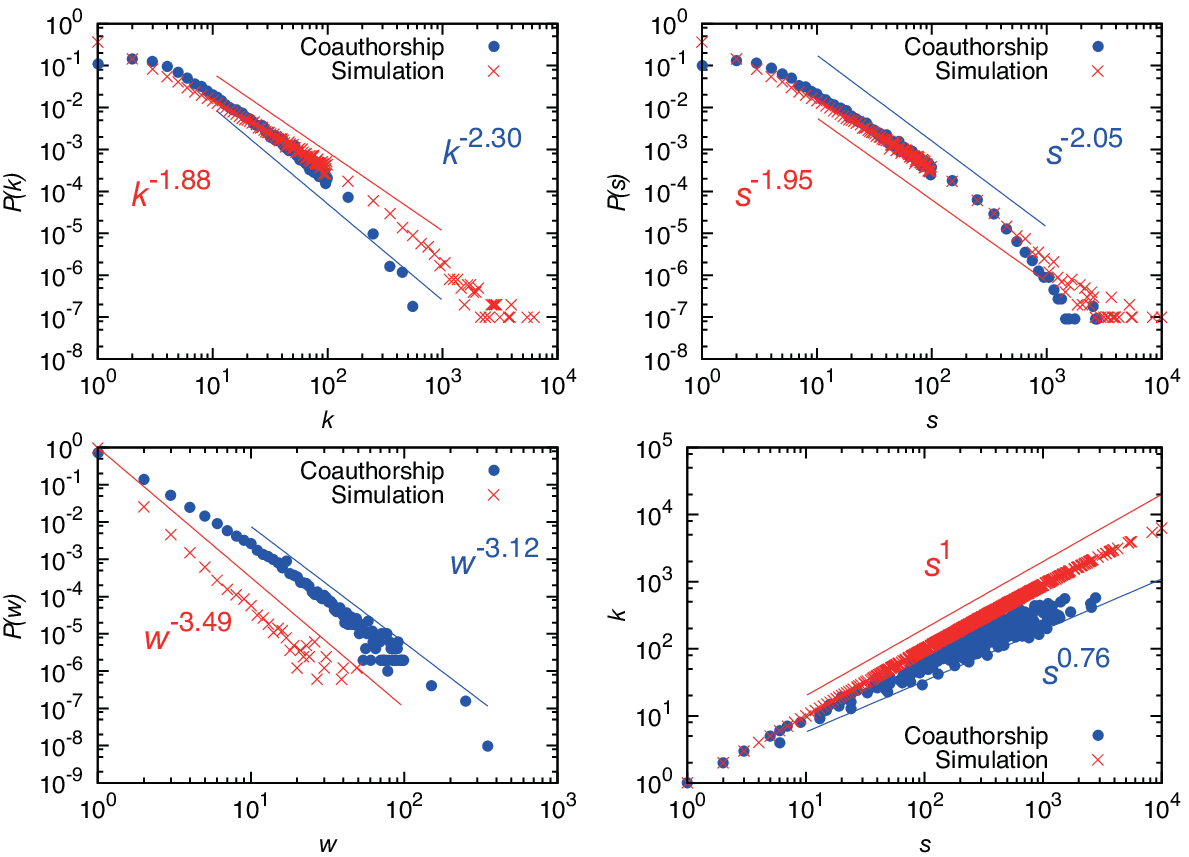}}
\caption{\label{geofig}
}
\end{figure}

The model of the weighted scale-free network we have presented in this
paper is simple enough to be analytically tractable, which enables us to
easily derive the statistical properties. In particular, only a single
control parameter $c$ determines the network properties, such as the
distributions of degree, strength and weight.
\changedtext{
In the coauthorship network of the researchers, the quantity $c$ can
be estimated as
$c=1.5\times 10^{-4}$ by the condition that, in the real data,  $t$ is the
number of researchers (100945) and $2t+ct^2$ must be equal to the
summation of the strength, $1.75 \times 10^6$.
The smallness of $c$ implies that new papers on which no new researcher collaborates, is rare.
The network generated by this model with the above estimated $c$ exhibits 
scale-free properties similar to the real coauthorship network of 
researchers (Fig.~\ref{geofig}).
The important point is that the various scale-free
properties and the exponents stem from a single real-measured parameter $c$.
The actor/actress collaboration network can be fitted quite well using the present
model (data not shown), and this network also has small $c$.
}
On the other hand, the
e-mail network can be regarded as a typical case of the large $c$, because
an enormous number of e-mails are communicated everyday, regardless of
whether new persons begin to use e-mail or not.
This means that the
weight of the link between the existing nodes tends to increase
independently of the addition of new nodes.
The exponent of $P(k)$ the of e-mail network is reported to be around 1.8 \cite{Ebel:2002}, 
which lies between the exponents 3/2 and 2 of our model network with
large $c$.

In conclusion, we have proposed a novel type of weighted scale-free
network model.
The significant, novel rule in this model is that the
weight of the links associated with some constant fraction of existing nodes (represented by the
parameter $c$) spontaneously increases independently of the attachment of
new nodes.
As a result, two types of scale-free network emerge depending
on the parameter $c$.
The resultant networks for small $c$ and large $c$ seem
to capture the statistical properties of the 
coauthorship network and e-mail network, respectively.
This suggests that the proposed simple algorithm is
suitable for studying certain types of real-world social networks.
Another important point is its analytical tractability, which means that some statistical properties 
can be derived theoretically in this model. 
This is helpful not only for a deeper understanding of the weighted scale-free 
networks, but also for developing some extended version of the model, and thus for studies 
in the other fields related to complex systems, such as oscillators \cite{Chavez:2005}, 
epidemics \cite{Hufnagel:2004}, and biological networks \cite{Jeong:2001}.
We believe that this model provides some insights on the dynamical evolution of such social
networks, leading to the understanding of more general mechanisms
underlying complex networks.

\begin{acknowledgments}
This work was supported by Grants-in-Aid from the Ministry of Education, Science, Sports, and
 Culture of Japan: Grant numbers 18047014,  18019019, and 18300079.
\end{acknowledgments}

\newpage


\newpage

Fig. \ref{schema}\\
At each time step, a new single node (a blue circle) appears and
connects to one existing node with a link of weight one (a blue link).
This new link is created by preferential attachment with the
probability proportional to the strength of the existing node. 
At the same time,
some pairs of existing nodes are chosen on a simple strength preferential rule 
(see the main text for details), and
the weights of the links between these chosen nodes (a green link) increase by one.
If no corresponding link exists, a new link of weight one (a red link) is created.
The numbers on the nodes and near the links indicate the strengths and the
weights, respectively.

Fig. \ref{fig1}\\
Comparison of distribution of strength between theoretical and
numerical results.
Note that for the network with $c=1.75\times 10^{-4}$ the power law exponent
changes at the crossover point indicated by the arrow.
The bin width is set to 1 for $s< 100$ and 100 for $s> 100$ because points are
 sparse in the region $s> 100$.

Fig. \ref{fig3}\\
Distribution of degree. The theoretical result is that $P(k)\approx\sqrt{3/8}k^{-3/2} (k<k_c)$ and $\sqrt{\pi /c}k^{-2} (k>k_c)$
for the large $c$ network $c=0.5$ ($k_c=3(ct)^2/2^{5/3}$). 
For the small $c$ network $c=1.75\times 10^{-4}$, $P(k)\approx\sqrt{3/8}k^{-3/2}$ $(k<k_c)$ and 
$9(ct)^3/(16k^3)$ $(k>k_c)$ ($k_c=8\pi/(3c)$). 
Each point of crossover $k_c$ is indicated by an arrow.

Fig. \ref{fig5}\\
Distribution of weights.
No crossover behavior is observed, because $cu\ll 1$ ($cu\gg 1$) holds for almost all nodes 
in the network $c=1.75\times 10^{-4}$ ($c=0.5$).

Fig. \ref{geofig}\\
Comparison of scale-free properties between the 
coauthorship network (filled circles) and the present model (cross).
Degree distribution (top left), strength distribution (top right), weight
distribution (bottom left), and strength-degree relationship (bottom
right) are shown. 
The coauthorship network is reconstructed from the Geological 
 Literature Search System (GEOLIS+ CD-ROM Ver.5) provided by 
AIST (permission number 63500-A-20070322-001). 

\newpage

\end{document}